\documentclass{article}
\usepackage{epsfig}

\begin{document}
\begin{large}
\pagenumbering{arabic}

\begin{center}
\bigskip
{\large\bf Inelastic $J/\psi$ production at HERA\\
in the colour singlet model with $k_T$-factorization}\\

\vspace{1.5cm}

{\large 
A.V.~Lipatov\footnote{E-mail: artem\_lipatov@mail.ru}\,

{\it Physical Department, M.V. Lomonosov Moscow State University,\\
119992 Moscow, Russia\/}\\[3mm]}

\vskip 0.5cm

N.P.~Zotov\footnote{E-mail: zotov@theory.sinp.msu.ru}

{\it D.V.~Skobeltsyn Institute of Nuclear Physics,\\ 
M.V. Lomonosov Moscow State University,\\
119992 Moscow, Russia\/}\\[3mm]

\end{center}

\vspace{1.0cm}

\begin{center}
{\bf{Abstract}}
\end{center}
\bigskip

In the framework of $k_T$-factorization QCD approach and the colour 
singlet model we consider $J/\psi$ inelastic photo- and leptoproduction 
processes at HERA. We investigate the dependences of the single differential 
and double differential cross section on different forms of the 
unintegrated gluon distribution. The $z$ and ${\bf p}_T$ dependences of the 
spin aligment parameter $\alpha$ are presented also.
Our theoretical predictions agree well with recent data taken by the
H1 and ZEUS collaborations at HERA. It is shown that experimental
study of the polarization $J/\psi$ mesons at low $Q^2 < 1\,{\rm GeV}^2$ is 
an additional test of BFKL gluon dynamics.

\vspace{1.5cm}

\section{Introduction} \indent 

It is known that from heavy quark and quarkonium production processes 
one can obtain unique information on gluon structure function of the 
proton because of the dominance of the photon-gluon or gluon-gluon fusion 
subprocess in the framework of QCD~[1]. Studying gluon distributions 
at modern collider energy (such as HERA, Tevatron) is important for 
prediction of heavy quark and quarkonium production cross sections 
at future colliders (LHC, THERA). At the energies of HERA and LEP/LHC 
colliders heavy quark and quarkonium production processes are so called 
semihard processes~[2--5]. In such processes by definition the hard scattering 
scale $\mu \sim m_Q$ is large compare to the $\Lambda_{{\rm QCD}}$ parameter 
but on the other hand $\mu$ is much less than the total center-of-mass 
energy: $\Lambda_{{\rm QCD}}\ll \mu\ll\sqrt s$. The last condition 
implies that the processes occur in small $x$ region: 
$x\simeq m_Q/\sqrt s\ll 1$, and that the cross sections of heavy quark 
and quarkonium production processes are determined by the behavior of 
gluon distributions in the small $x$ region.

It is also known that in the small $x$ region the standard parton 
model (SPM) assumptions about factorization of gluon distribution functions 
and subprocess cross sections are broken because the subprocess cross sections 
and gluon structure functions depend on a gluon transverse momentum 
$k_T$~[2--5]. So calculations of heavy quark production cross sections at 
HERA, Tevatron, LHC and other collider conditions are necessary to carry 
out in the so called $k_T$-factorization (or semihard) QCD approach, which 
is more preferable for the small $x$ region than SPM.

The $k_T$-factorization QCD approach is based on Balitsky, Fadin, Kuraev, 
Lipatov (BFKL)~[6] evolution equations. The resummation of the terms 
$\alpha_{S}^n\,\ln^n(\mu^2/\Lambda_{{\rm QCD}}^2)$, 
$\alpha_{S}^n\,\ln^n(\mu^2/\Lambda_{{\rm QCD}}^2)\,\ln^n(1/x)$ and 
$\alpha_{S}^n\,\ln^n(1/x)$ in the $k_T$-factorization approach leads to the 
unintegrated (dependent from ${\bf q}_T$) gluon distribution 
$\Phi(x,{\bf q}_T^2,\mu^2)$ which determine the probability to find a gluon 
carrying the longitudinal momentum fraction $x$ and transverse momentum 
${\bf q}_T$ at probing scale $\mu^2$.

To calculate the cross section of a physical process the unintegrated gluon
distributions have to be convoluted with off mass shell matrix elements
corresponding to the relevant partonic subprocesses~[2--5]. In the off mass
shell matrix element the virtual gluon polarization tensor is taken in
the BFKL form~[2--5]:
\begin{equation}
L^{\mu\,\nu}(q) = {q_T^{\mu}\,q_T^{\nu}\over {\bf q}_T^2}.
\end{equation}

Nowadays, the significance of the $k_T$-factorization QCD approach
becomes more and more commonly recognized~[7]. It was already used for the 
description of a wide class heavy quark and quarkonium production 
processes~[8--23]. It is notable that calculations in $k_T$-factorization 
approach provide results which are absent in other approaches, such as the 
fast growth of total cross sections in comparison with SPM, a broadening of 
the $p_T$ spectra due to extra the transverse momentum of the colliding 
partons and other polarization properties of final particles in comparison 
with SPM.

We point out that heavy quark and quarkonium cross section calculations 
within the SPM in the fixed order of pQCD have some problems. For example, 
the very large discrepancy (by more than an order of magnitude)~[24, 25] 
between the pQCD predictions for hadroproduction $J/\psi$ and $\Upsilon$ 
mesons and experimental data at Tevatron was found. This fact has resulted 
in intensive theoretical investigations of such processes. In particular, 
it was required to use additional transition mechanism from $c\bar c$-pair 
to the $J/\psi$ mesons, so-called the colour octet (CO) model~[26], where 
$c\bar c$-pair is produced in the color octet state and transforms into 
final colour singlet (CS) state by help soft gluon radiation. The CO model 
was supposed to be applicable to heavy quarkonium hadro- and leptoproduction 
processes. However, the contributions from the CO mechanism to the $J/\psi$ 
meson photoproduction contradict the experimental data at HERA for 
$z$ distribution~[27--30].

Another difficulty of the CO model are the $J/\psi$ polarization properties 
in $p\bar p$-interactions at the Tevatron. In the framework of the CO model, 
the $J/\psi$ mesons should be transverse polarized at the large transverse 
momenta ${\bf p}_T$. However, this is in contradiction with the experimental 
data, too.

The CO model has been applied earlier~[31, 32] in an analysis of the $J/\psi$ 
inelastic production experimental data at HERA~[33]. However, 
the results do not agree with each other~[32]. We note that the 
shapes of the $Q^2$, ${\bf p}_T^2$ and $y^*$ distributions are not
reproduced by the calculation~[31], and the $z$ distributions~[32] contradict 
the HERA experimental data too. Results obtained within the usual collinear 
approach and CS model~[34--37] underestimate experimental 
data by factor about 2. 

The inelastic $J/\psi$ production at HERA in the CS model with 
$k_T$-factorization also was considered in~[15, 16]. The results~[16] agree 
with H1 experimental data~[33] both in normalization and shape only at 
quite small charmed quark mass $m_c = 1.4\,{\rm GeV}$. The theoretical 
prediction~[15] are stimulated the experimental analysis of $J/\psi$ 
polarization properties at HERA conditions. 

Recently the new experimental data on the inelastic 
$J/\psi$ photo- and leptoproduction at HERA were obtained by the 
H1~[38, 39] and ZEUS~[40] collaborations with increased statistics 
and precision as compared with previous experimental 
analyses~[33]. Based on the above mentioned results here we will use the 
CS model and the $k_T$-factorization approach for the analysis of the 
data~[38--40]. We investigate the dependences of the single 
differential and double differential $J/\psi$ production cross section on 
different forms of the unintegrated gluon distribution. Special attention is 
drawn to the unintegrated gluon distributions obtained from BFKL evolution
equation which has been applied earlier in our previous papers~[12--16]. 
For studying $J/\psi$ meson polarization properties we calculate the 
$z$ and ${\bf p}_T$ dependences of the spin aligment parameter $\alpha$. 

The outline of this paper is as follows. In Section 2 we present, in 
analytic form, the total and differential cross section for the
inelastic $J/\psi$ photo- and leptoproduction in the CS model with 
$k_T$-factorization, and give the formulas for the relevant partonic 
subprocess off mass shell matrix elements. In Section 3 we present the 
numerical results of our calculations and compare them with the
H1~[38, 39] and ZEUS~[40] data. Finally, in Section 4, we give some 
conclusions.

\bigskip

\section{Analytic results} \indent 

In this section we calculate total and differential cross section for 
inelastic $J/\psi$ photo- and leptoproduction in the CS model with 
$k_T$-factorization, and give the formulas for the relevant partonic subprocess 
off mass shell matrix elements.

\subsection{Kinematics} \indent 

As indicated in Fig.~1, we denote the 4-momenta of the incoming electron and
proton and the outgoing electron, proton remnant, $J/\psi$ meson and  
gluon by $p_e$, $p_p$, $p_e^{\prime}$, $p_p^{\prime}$, $p_{\psi}$ and 
$p_g$, respectively. The initial virtual photon and BFKL gluon have a 
4-momenta $q_1 = p_e - p_e^{\prime}$ and $q_2 = p_p - p_p^{\prime}$, so that 
4-momentum transfer $Q^2 = - q_1^2$. In our analysis below we will use the 
Sudakov decomposition, which has the following form:
\begin{equation}
p_{\psi} = \alpha_1 p_e + \beta_1 p_p + p_{\psi\,T},\qquad p_g = \alpha_2 p_e + \beta_2 p_p + p_{g\,T},\atop
\quad {\quad {q_1 = x_1 p_e + q_{1T},\qquad q_2 = x_2 p_p + q_{2T}}},
\end{equation}

\noindent where $p_{\psi\,T}$, $p_{g\,T}$, $q_{1\,T}$ and $q_{2\,T}$ are 
transverse 4-momenta of corresponding particles, and 
\begin{equation}
p_{\psi}^2 = m_{\psi}^2,\quad p_g^2 = 0,\quad q_1^2 = q_{1T}^2,\quad q_2^2 = q_{2T}^2.
\end{equation}

\noindent In the $ep$ c.m. frame we can write:
\begin{equation}
p_e = \sqrt s/2\,(1,\,0,\,0,\,1),\qquad p_p = \sqrt s/2\,(1,\,0,\,0,\,-1),
\end{equation}

\noindent where we neglect the masses of the electron and proton. The Sudakov 
variables are expressed as follows:
\begin{equation}
\displaystyle \alpha_1={m_{\psi\,T}\over {\sqrt s}}\exp(y_{\psi}),\qquad \alpha_2={|{\bf p}_{g\,T}|\over {\sqrt s}}\exp(y_g),\atop
\displaystyle \beta_1={m_{\psi\,T}\over {\sqrt s}}\exp(-y_{\psi}),\qquad \beta_2={|{\bf p}_{g\,T}|\over {\sqrt s}}\exp(-y_g),
\end{equation} 

\noindent where $m_{\psi\,T}^2 = m_{\psi}^2 + {\bf p}_{\psi\,T}^2$, 
$y_{\psi}$ and $y_g$ are the rapidities of $J/\psi$ meson and final gluon 
respectively in the $ep$ c.m. frame. From conservation laws we can easy
obtain following conditions:
\begin{equation}
x_1 = \alpha_1 + \alpha_2,\qquad x_2 = \beta_1 + \beta_2,\qquad 
{\bf q}_{1T} + {\bf q}_{2T} = {\bf p}_{\psi\,T} + {\bf p}_{g\,T}.
\end{equation}

Also the variable $z = (p_{\psi} \cdot p_p)/(q_1 \cdot p_p)$ is used
for a description of quarkonium photo- and leptoproduction processes. In
the rest frame of the proton one has $z = E_{\psi}/E_{\gamma}$.

\subsection{Inelastic $J/\psi$ leptoproduction cross section} \indent 

In the $k_T$-factorization approach the differential cross section for 
inelastic $J/\psi$ leptoproduction may be written as:
\begin{equation}
\displaystyle d\sigma(e\,p\to e'\,J/\psi\,X) = {dx_2 \over x_2}\,\Phi(x_2,\,{\bf q}_{2T}^2,\,\mu^2)\,{d\phi_2\over 2\pi}\,d{\bf q}_{2T}^2\,d\hat \sigma(e\,g^*\to e'\,J/\psi\,g'),
\end{equation}

\noindent where $\phi_2$ is initial BFKL gluon azimuthal angle, 
$\Phi(x_2,{\bf q}_{2T}^2,\mu^2)$ is an unintegrated gluon distribution in 
the proton. The $e\,g^*\to e'\,J/\psi\,g'$ cross section is given by: 
\begin{equation}
d\hat \sigma(e\,g^*\to e'\,J/\psi\,g') = \displaystyle {(2\pi)^4\over 2 x_2 s}\,\sum {|M|^2_{{\rm SHA}}(e\,g^*\to e'\,J/\psi\,g')}\,\times \atop
\displaystyle \times {d^3p'_e\over (2\pi)^3\,2{p'}_e^0}\,{d^3p_{\psi}\over (2\pi)^3\,2p_{\psi}^0}\,{d^3p_g\over (2\pi)^3\,2p_g^0}\,\delta^{(4)}(p_e + q_2 - p_e' - p_{\psi} - p_g),
\end{equation}

\noindent where $\sum {|M|^2_{{\rm SHA}}(e\,g^*\to e'\,J/\psi\,g')}$ is the 
off mass shell matrix element. In (8) $\sum$ indicates an averaging over and 
a sum over the final polarization states. From (7) and (8) we obtain the 
following formula for the inelastic $J/\psi$ leptoproduction differential 
cross section in the $k_T$-factorization approach:
\begin{equation}
\displaystyle d\sigma(e\,p\to e'\,J/\psi\,X) = {1\over 128\pi^3}\, {\Phi(x_2,\,{\bf q}_{2T}^2,\,\mu^2)\over (x_2\,s)^2\,(1 - x_1)}\,{dz\over z\,(1 - z)}\,dy_{\psi}\,\times \atop
\displaystyle \times\, \sum {|M|^2_{{\rm SHA}}(e\,g^*\to e'\,J/\psi\,g')}\,d{\bf p}_{\psi\,T}^2\,dQ^2\,d{\bf q}_{2T}^2\,{d\phi_1\over 2\pi}\,{d\phi_2\over 2\pi}\,{d\phi_{\psi}\over 2\pi},\end{equation}

\noindent where $\phi_1$ and $\phi_{\psi}$ are azimuthal angles of the 
initial virtual photon and $J/\psi$ meson respectively.

\subsection{Inelastic $J/\psi$ photoproduction cross section} \indent 

As in leptoproduction case, in the $k_T$-factorization approach the 
differential cross section for inelastic $J/\psi$ photoproduction may be 
written as:
\begin{equation}
\displaystyle d\sigma(\gamma\,p\to J/\psi\,X) = {dx_2 \over x_2}\,\Phi(x_2,\,{\bf q}_{2T}^2,\,\mu^2)\,{d\phi_2\over 2\pi}\,d{\bf q}_{2T}^2\,d\hat \sigma(\gamma\,g^*\to J/\psi\,g').
\end{equation}

\noindent If we take the limit $Q^2\to 0$ and $x_1 \to 1$, 
we easy obtain the following formula for the inelastic $J/\psi$ 
photoproduction differential cross section in the $k_T$-factorization 
approach by analogy with the leptoproduction case: 
\begin{equation}
\displaystyle d\sigma(\gamma\,p\to J/\psi\,X) = {1 \over 16\pi\,(x_2\,s)^2}\,\Phi(x_2,\,{\bf q}_{2T}^2,\,\mu^2)\,{dz\over z\,(1 - z)}\,\times \atop
\displaystyle \times\, \sum {|M|^2_{{\rm SHA}}(\gamma\,g^*\to J/\psi\,g')}\,d{\bf p}_{\psi\,T}^2\,d{\bf q}_{2T}^2\,{d\phi_2\over 2\pi}\,{d\phi_{\psi}\over 2\pi},
\end{equation}

We note that formulas for the the differential cross section for 
inelastic $J/\psi$ photo- and leptoproduction in the usual parton model
may be obtained from (9) and (11), if we take the limit ${\bf q}_{2T}^2\to 0$ 
and average them over the transverse momentum vector ${\bf q}_{2T}$.

\subsection{Off mass shell matrix element} \indent 

There are six Feynman diagrams (Fig.~2) which describe partonic subprocess
$\gamma\,g^*\to J/\psi\,g'$ at leading order in $\alpha_{S}$ and $\alpha$. 
In the framework of the CS model and the nonrelativistic approximation 
the production of the $J/\psi$ meson is considered as a production of a 
quark-antiquark system in the colour singlet state with orbital momentum
$L = 0$ and spin momentum $S = 1$. The binding energy and relative 
momentum of the quarks in the $J/\psi$ meson are neglected, resulting in 
$m_{\psi} = 2\,m_c$, where $m_c$ is charm mass. The amplitude of the 
process $\gamma\,g^*\to J/\psi\,g'$ may be obtained from the amplitude of the 
process $\gamma\,g^*\to c\bar c\,g'$ after replacement:
\begin{equation}v(p_{\bar c})\,\bar u(p_c) \to \hat J(p_{\psi}) = {\psi(0)\over 2\sqrt {\mathstrut m_{\psi}}}\,\hat \epsilon(p_{\psi})\,(\hat p_{\psi} + m_{\psi})\,{1\over \sqrt {\mathstrut 3}},
\end{equation}

\noindent where $p_c = p_{\psi}/2$, $\epsilon(p_{\psi})$ is a 4-vector of 
the $J/\psi$ polarization, $1/\sqrt 3$ is the color factor, $\psi(0)$ is the 
nonrelativistic meson wave function at the origin. The matrix element is:
\begin{equation}
\displaystyle M = e_c\,g^2\,\epsilon_{\mu}(q_1)\,\epsilon_{\sigma}(q_2)\,\epsilon_{\rho}(p_g)\,\times \atop 
\displaystyle \times Sp \left[\hat J(p_{\psi})\,\gamma^{\mu}\,{\hat p_{c} - \hat q_1 + m_c\over (p_c - q_1)^2 - m_c^2}\,\gamma^{\sigma}\,{ - \hat p_c - \hat p_g + m_c\over (- p_c - p_g)^2 - m_c^2}\,\gamma^{\rho}\right]
\end{equation}

\noindent + 5 permutations of all gauge bosons. Here $\epsilon_{\mu}(q_1)$ and 
$\epsilon_{\mu}(q_2)$ are polarization vectors of the initial photon and 
gluon respectively, $\epsilon_{\mu}(p_g)$ is a 4-vector of the final gluon 
polarization. The summation on the $J/\psi$ meson and final gluon 
polarizations is carried out by covariant formulas:
\begin{equation}
\sum {\epsilon^{\mu}(p_{\psi})\epsilon^{*\,\nu}(p_{\psi})} = - g^{\mu\nu} + {p_{\psi}^{\mu}\,p_{\psi}^{\nu}\over m_{\psi}^2},
\end{equation}
\begin{equation}
\sum {\epsilon^{\mu}(p_g)\epsilon^{*\,\nu}(p_g)} = - g^{\mu\nu}.
\end{equation}

\noindent The initial BFKL gluon polarization tensor is taken in form (1). 
For the photon we use the usual expression
\begin{equation}
\sum {\epsilon^{\mu}(q_1)\epsilon^{*\,\nu}(q_1)} = - g^{\mu\nu}
\end{equation}

\noindent in photoproduction case and the full lepton tensor 
(including also the photon propagator factor and photon-lepton coupling) in
leptoproduction case:
\begin{equation}
\sum {\epsilon^{\mu}(q_1)\epsilon^{*\,\nu}(q_1)} = 2\,{e^2\over Q^2}\,\left( - g^{\mu\nu} + {4 p_{e}^{\mu}p_{e}^{\nu}\over Q^2}\right).
\end{equation}

For studying $J/\psi$ polarized production we introduce the 
4-vector of the longitudinal polarization $\epsilon_L^{\mu}(p_{\psi})$ as 
follows~[41]:
\begin{equation}
\epsilon^{\mu}_L(p_{\psi}) = {(p_{\psi}\cdot p_p)\over \sqrt {\mathstrut {(p_{\psi}\cdot p_p)^2 - m_{\psi}^2}\,s}}\left({p_{\psi}^{\mu} \over m_{\psi}} - {m_{\psi}p_p^{\nu}\over (p_{\psi}\cdot p_p)}\right).
\end{equation}

The evaluation of $\sum{|M|_{{\rm SHA}}^2}$ for photo- and leptoproduction
cases was done analytically by the REDUCE program. Also in our calculations 
we have used the JB~[42] and KMS~[43] parametrizations of the unintegrated gluon 
distributions (see also~[7] for the detail information).

\bigskip

\section{Numerical results} \indent 

In this section we present the theoretical results in comparison with 
recent experimental data taken by the H1~[38, 39] and ZEUS~[40] 
collaborations at HERA.

There are three parameters which determine the common normalization factor 
of the cross section under consideration: $J/\psi$ meson wave function at 
the origin $\psi(0)$, charmed quark mass $m_c$ and factorization scale $\mu$. 
The value of the $J/\psi$ meson wave function at the origin may be calculated 
in a potential model or obtained from the well known experimental decay 
width $\Gamma(J/\psi \to \mu^{+}\,\mu^{-})$. In our calculation we 
used $|\psi(0)|^2 = 0.0876\,{\rm GeV}^3$ as in~[44].

Concerning a charmed quark mass, the situation is not clear: on the one hand, 
in the nonrelativistic approximation one has 
$m_c = m_{\psi}/2 = 1.55\,{\rm GeV}$, but on the other hand there are
examples when smaller value of a charm mass $m_c = 1.4\,{\rm GeV}$ is 
used~[32, 45]. However, in our previous paper~[16] we analyzed in detail
the influence of charm quark mass on the theoretical results.
We found that the main effect of change of the charm quark mass connects with 
final phase space of $J/\psi$ meson, and in the subprocess matrix elements 
this effect is neglectable. Taking into account that the value of 
$m_c = 1.4\,{\rm GeV}$ corresponds to the unphysical phase space of 
$J/\psi$ state, in the present paper we will use value of a charm 
mass $m_c = 1.55\,{\rm GeV}$ only.

Also the most significant theoretical uncertanties come from the choice of 
the factorization scale $\mu_F$ and renormalization one $\mu_R$. One of them 
is related to the evolution of the gluon distributions 
$\Phi(x,{\bf q}_T^2,\mu_F^2)$, the other is responsible for strong coupling 
constant $\alpha_{S}(\mu_R^2)$. As often done in literature, we set 
$\mu_F = \mu_R = \mu$. In the present paper we used the following 
choice $\mu^2 = {\bf q}_{2T}^2$ as in~[16, 46].

\subsection{Inelastic $J/\psi$ leptoproduction at HERA} \indent 

The integration limits in (9) are taken as given by kinematical conditions 
of the H1 experimental data~[39]. One kinematical region\footnote{Here
we denote the $J/\psi$ meson transverse momentum and 
rapidity in the $\gamma^*\,p$ c.m. frame by ${\bf p}_{\psi\,T}^{*}$ and
$y_{\psi}^{*}$, respectively.} is $2 < Q^2 < 100\,{\rm GeV}^2$, $50 < W < 225 \,{\rm GeV}$, 
$0.3 < z < 0.9$, ${\bf p}_{\psi\,T}^{*\,2} > 1\,{\rm GeV}^2$ and other 
kinematical region is $12 < Q^2 < 100\,{\rm GeV}^2$, 
$50 < W < 180 \,{\rm GeV}$, ${\bf p}_{\psi\,T}^{2} > 6.4\,{\rm GeV}^2$, 
$0.3 < z < 0.9$ and ${\bf p}_{\psi\,T}^{*\,2} > 1\,{\rm GeV}^2$.
Here and in the following, we used $\Lambda_{{\rm QCD}} = 250\,{\rm MeV}$.

The results of our calculations are shown in Fig.~3---5. Fig.~3 shows the 
single differential cross sections of the inelastic $J/\psi$ meson 
leptoproduction obtained in the first kinematical region 
at $\sqrt s = 314\,{\rm GeV}$. 
Curve {\sl 1} corresponds to the SPM calculations at the leading order 
approximation with the GRV (LO) gluon density, curves {\sl 2} and {\sl 3} 
correspond to the $k_T$-factorization 
results with the JB (at $\Delta = 0.35$~[17, 23]) and the KMS unintegrated gluon 
distributions. One can see that results obtained in the CS model with 
$k_T$-factorization agree very well with the H1 experimental data. 
The SPM calculation are lower than the data by a factor 2 --- 3.

We would like to note the difference in the transverse momenta
distribution shapes between curves obtained using the $k_T$-factorization 
approach and the SPM. This difference manifests the $p_T$ broadening effect which 
mentioned earlier. It is visible also that only the $k_T$-factorization 
approach gives a correct description of the ${\bf p}_{\psi\,T}^{2}$ spectra. 
However, we note that the ${\bf p}_{\psi\,T}^{*\,2}$ distributions 
somewhat less well described (in contrast with ${\bf p}_{\psi\,T}^{2}$ 
spectra) at the large values of the $J/\psi$ transverse momenta (see Fig.~3d).

Also we point out the good description of the $z$ distributions 
which obtained in the $k_T$-factorization approach in contrast with 
CO model results~[32], except for the region $z < 0.3$, where the 
contribution of the resolved photon process may be large~[47].

Fig.~4 shows the single differential cross sections of the inelastic 
$J/\psi$ meson production obtained in the second kinematical region at 
$\sqrt s = 314\,{\rm GeV}$. Curves {\sl 1 --- 3} are the same 
as in Fig.~3. We find also good agreement between results obtained 
in the CS model with $k_T$-factorization and H1 data.
It is notable that in this kinematical region in contrast with first one 
the both ${\bf p}_{\psi\,T}^{2}$ and ${\bf p}_{\psi\,T}^{*\,2}$ transverse 
momenta distributions agree well with the experimental data.

The double differential cross sections $d\sigma/dQ^2\,dz$ and 
$d\sigma/d{\bf p}_{\psi\,T}^{*\,2}\,dz$ (Fig.~5) obtained 
with $k_T$-factorization in the different $z$ 
regions $0.3 < z < 0.6$ (Fig.~5a, b), $0.6 < z < 0.75$ (Fig.~5c, d) 
and $0.75 < z < 0.9$ (Fig.~5e, f) agree with the H1 data. We note that 
double differential cross sections $d\sigma/d{\bf p}_{\psi\,T}^{*\,2}\,dz$ somewhat less 
well described at the large $z$ (see Fig.~5f).
However, in this region the contribution of the diffractive processes 
may be large. All of these contributions are not in our consideration.

It is interesting to note that results obtained with the JB unintegrated gluon 
distribution at $\Delta = 0.35$ and the KMS ones, which effectively included
about 70\% of the full NLO corrections to the value of $\Delta$~[43], coincide
practically in a wide kinematical region. 

Fig.~3 --- 5 show that the $k_T$-factorization results for inelastic 
$J/\psi$ leptoproduction with realistic value of a charm  
mass $m_c = 1.55\,{\rm GeV}$ agree well with the
H1 experimental data without any additional 
$c\bar c \to J/\psi$ fragmentation mechanisms, such as the CO contributions.

\subsection{Inelastic $J/\psi$ photoproduction at HERA} \indent 

The integration limits in (11) are taken as given by kinematical conditions 
of the H1~[38] and the ZEUS~[40] data. One kinematical 
region which corresponds to the H1 experiment is $60 < W < 240 \,{\rm GeV}$, 
$0.3 < z < 0.9$, $1 < {\bf p}_{\psi\,T}^{2} < 60\,{\rm GeV}^2$ and other 
kinematical region which corresponds to the ZEUS experiment is 
$50 < W < 180 \,{\rm GeV}$, $0.4 < z < 0.9$, 
${\bf p}_{\psi\,T}^{2} > 1\,{\rm GeV}^2$.

The results of our calculations are shown in Fig.~6 --- 8. Fig.~6 and 7 show 
the total and single differential cross sections of inelastic 
$J/\psi$ meson photoproduction in comparison with the H1 and ZEUS data, 
respectively. As in previous section, curve {\sl 1} corresponds to the SPM 
calculations at the leading order approximation with the GRV (LO) gluon 
density, curves {\sl 2} and {\sl 3} correspond to the $k_T$-factorization 
results with the JB (at $\Delta = 0.35$~[17, 23]) and KMS unintegrated gluon 
distributions.

The $W$ dependences of the total $J/\psi$ photoproduction 
cross section at $0.3 < z < 0.9$, $0.3 < z < 0.8$ and $0.4 < z < 0.9$ are 
plotted in Fig.~6a, Fig.~6b and Fig.~7a respectively. One can see that 
results obtained in the CS model with $k_T$-factorization agree very well 
with the H1~[38] and the ZEUS~[40] experimental data. The SPM results 
are lower than the data by a factor 2.

Concerning the shapes of the ${\bf p}_{\psi\,T}^{2}$ distribution (Fig.~6c and
7b), one can note a difference between the $k_T$-factorization and the SPM 
curves. As in leptoproduction case, this difference 
manifests the $p_T$ broadening effect which was mentioned earlier. It is 
visible also that only the $k_T$-factorization approach gives a 
correct description of the H1 data.

The $z$ distributions are shown in Fig.~6d, e, f and Fig.~7c, d, e at 
different ${\bf p}_{\psi\,T}$ cuts in comparison with the H1 and ZEUS data, 
respectively. One can see 
that good agreement between the $k_T$-factorization 
curves and the experimental data is observed. The $z$ distribution somewhat 
are less well described at ${\bf p}_{\psi\,T} > 3\,{\rm GeV}$ (see Fig.~6f).
The disperance between the leading order SPM calculations and the experimental 
data is about factor 2 at ${\bf p}_{\psi\,T} > 1\,{\rm GeV}$ and
about order of magnitude at ${\bf p}_{\psi\,T} > 3\,{\rm GeV}$ 
for $z \sim 0.8$. Also we note that in the region $z < 0.3$ the contribution 
of resolved photon process may be large~[47], as in leptoproduction case.

The double differential cross sections $d\sigma/d{\bf p}_{\psi\,T}^2\,dz$
(Fig.~8) in the different $z$ regions $0.3 < z < 0.6$ (Fig.~8a),
$0.6 < z < 0.75$ (Fig.~8b) and $0.75 < z < 0.9$ (Fig.~8c) are well described by 
the $k_T$-factorization approach.

It can be seen that the results obtained with the JB unintegrated gluon distribution
with $\Delta = 0.35$ 
and the KMS ones (which effectively included the main part of the full NLO 
corrections to the value of $\Delta$) practically coincide in a wide kinematical region,
as in leptoproduction case. 

Fig.~6 --- 8 show that the $k_T$-factorization results for inelastic 
$J/\psi$ photoproduction with realistic value of a charm  
mass $m_c = 1.55\,{\rm GeV}$ agree well with 
the H1 and ZEUS experimental data without any additional $c\bar c \to J/\psi$ 
fragmentation mechanisms, such as CO contributions.

\subsection{Polarization properties of the $J/\psi$ meson at HERA} \indent 

As it mentioned above, one of differences between the $k_T$-factorization 
approach and the SPM is connected with polarization properties of the 
final particles. In the present paper for studying $J/\psi$ meson 
polarization properties we calculate the ${\bf p}_T$ and $z$ dependences 
of the spin aligment parameter $\alpha$~[14--16]:
\begin{equation}
\alpha (\omega) = {d\sigma/d\omega - 3\,d\sigma_L/d\omega\over d\sigma/d\omega + d\sigma_L/d\omega},
\end{equation}

\noindent where $\sigma_L$ is the production cross section for the 
longitudinally polarized $J/\psi$ mesons, $\omega = {\bf p}_{\psi\,T},\,z$.
The parameter $\alpha$ controls the angular distribution for leptons in the 
decay $J/\psi \to \mu^{+}\,\mu^{-}$ (in the $J/\psi$ meson rest frame):
\begin{equation}
{d\Gamma(J/\psi \to \mu^{+}\,\mu^{-})\over d\cos\theta} \sim 1 + \alpha\,\cos^2\theta.
\end{equation}

\noindent The cases $\alpha = 1$ and $\alpha = - 1$ correspond to transverse
and longitudinal polarization of the $J/\psi$ meson, respectively. 

In our previous paper~[16] we analyzed in detail the $Q^2$ and 
${\bf p}_{\psi\,T}^2$ dependences of the spin parameter $\alpha$ in 
leptoproduction case. We found that it is impossible to make of exact 
conclusions about a BFKL gluon contribution to the polarized 
$J/\psi$ production cross section because of large additional 
contribution from initial longitudinal polarization of virtual photons.
However at low $Q^2$ and in photoproduction limit 
these contributions are negligible. This fact should result in 
observable spin effects of final $J/\psi$ mesons, connected with the 
$k_T$-factorization effects. In this paper we have 
performed such calculations for the inelastic 
$J/\psi$ photoproduction process. 

The results of our calculations are shown in Fig.~9 and 10. Fig.~9 shows 
the parameter $\alpha$ as a function $z$ and 
${\bf p}_{\psi\,T}$ in comparison with the H1 experimental 
data which obtained in the kinematical region $60 < W < 240 \,{\rm GeV}$, 
$0.3 < z < 0.9$ and $1 < {\bf p}_{\psi\,T}^{2} < 60\,{\rm GeV}^2$. 
Curve {\sl 1} corresponds to the SPM calculations in the leading order 
approximation with the GRV (LO) gluon density, curve {\sl 2} corresponds 
to the $k_T$-factorization results obtained with the JB (at $\Delta = 0.35$~[17, 23]) 
unintegrated gluon distribution. One can see that the $z$ dependence of the spin parameter 
$\alpha$ is not sensitive to the results of different approaches, included the 
nonrelativistic QCD predictions (see [38]). However, the behavior of the 
$\alpha({\bf p}_{T})$ is different in the $k_T$-factorization approach and
the SPM (see Fig.~9b). Although the experimental points have large errors they
tends to support the $k_T$-factorization theoretical predictions.

Fig.~10 shows the ${\bf p}_{\psi\,T}$ dependence of the spin parameter 
$\alpha$ in comparison with the ZEUS experimental data which obtained in 
the kinematical region $50 < W < 180 \,{\rm GeV}$, $0.4 < z < 0.9$ 
(Fig.~10a and 10c) and $0.4 < z < 1$ (Fig.~10b and 10d). 
We note that in Fig.~10a and 10b the quantisation axis is chosen to be opposite
of the incoming proton direction in the $J/\psi$ rest frame, $\theta$ is
the opening angle between the quantisation axis and the $\mu^{+}$ direction
of flight in the $J/\psi$ rest frame. This frame is known as 
the "target frame"~[40]. In Fig.~10c and 10d, the quantisation axis was
defined as the $J/\psi$ direction of flight in the ZEUS coordinate system.
This frame is known as the "helicity basis"~[40, 48]. 
Curves {\sl 1} and {\sl 2} are the same as in Fig.~9.

It is visible that only the $k_T$-factorization approach gives a correct 
description of the ZEUS data, although the experimental points have 
large errors. We also have large difference between predictions of the 
leading order of SPM and the $k_T$-factorization approach.
The SPM predictions lies somewhat below the data at low ${\bf p}_{\psi\,T}$ 
and somewhat above at high ${\bf p}_{\psi\,T}$.
Therefore experimental measurement of polarization properties of 
the $J/\psi$ mesons will be an additional test of 
BFKL gluon dynamics.

\bigskip

\section{Conclusions} \indent 

In this paper we considered the inelastic $J/\psi$ meson photo- and leptoproduction 
at HERA
in the colour singlet model using the standard parton model in leading order 
in $\alpha_{S}$ and the $k_T$-factorization QCD approach. We investigated 
the total cross section, single differential and double differential 
cross sections of inelastic $J/\psi$ production on different forms of 
the unintegrated gluon 
distribution. The ${\bf p}_T$ and $z$ dependences of the spin aligment 
parameter $\alpha$ presented also. We compared the theoretical results with 
recent experimental data taken by the H1 and ZEUS collaboration at HERA. 
We have found that the $k_T$-factorization results (in contrast with the 
SPM ones) with the JB and KMS unintegrated gluon distributions agree well with the 
experimental data at realistic value of a charm mass $m_c = 1.55\,{\rm GeV}$, 
$|\psi(0)|^2 = 0.0876\,{\rm GeV}^3$ and $\Lambda_{{\rm QCD}} = 250\,{\rm MeV}$
without any additional transition mechanism from $c\bar c$-pair to the 
$J/\psi$ mesons (such as given by the CO model). 
We also found that results obtained with the JB unintegrated gluon 
density at $\Delta = 0.35$ and KMS one, which effectively included
about 70\% of the full NLO corrections to the Pomeron intercept $\Delta$, 
practically coincide in a wide kinematical region for $J/\psi$ production 
processes at HERA conditions. Finally, it is shown that experimental study 
of a polarization of $J/\psi$ meson at low $Q^2 < 1\,{\rm GeV}^2$ should 
be additional test of BFKL gluon dynamics.

\bigskip
\section{Acknowledgments} \indent 

The authors would like to thank S.~Baranov for encouraging interest and 
useful discussions. A.L. thanks also V.~Saleev for the help on the initial 
stage of work. The study was supported in part by RFBR grant 02--02--17513 
and INTAS grant YS 2002 N399.

\newpage
\thispagestyle{empty}

\noindent
{\bf Fig.~1.} Diagram for $e\,p\to e'\,J/\psi\,X$ process.

\vspace{1cm}
\noindent
{\bf Fig.~2.} Feynmans diagram used for description partonic process 
$\gamma\,g\to J/\psi\,g'$ process.

\vspace{1cm}
\noindent
{\bf Fig.~3.} The single differential cross sections of the inelastic 
$J/\psi$ leptoproduction obtained in the kinematical region 
$2 < Q^2 < 100\,{\rm GeV}^2$, $50 < W < 225 \,{\rm GeV}$, 
$0.3 < z < 0.9$ and ${\bf p}_{\psi\,T}^{*\,2} > 1\,{\rm GeV}^2$ at 
$\sqrt s = 314\,{\rm GeV}$  in comparison with the H1~[39] data. 
Curve {\sl 1} corresponds to the SPM calculations 
at the leading order approximation with GRV (LO) gluon density, 
curves {\sl 2} and {\sl 3} correspond to the $k_T$-factorization QCD 
calculations with JB and KMS unintegrated gluon distribution.

\vspace{1cm}
\noindent
{\bf Fig.~4.} The single differential cross sections of the inelastic 
$J/\psi$ leptoproduction obtained in the kinematical region 
$12 < Q^2 < 100\,{\rm GeV}^2$, $50 < W < 225 \,{\rm GeV}$, 
${\bf p}_{\psi\,T}^{2} > 6.4\,{\rm GeV}^2$,
$0.3 < z < 0.9$ and ${\bf p}_{\psi\,T}^{*\,2} > 1\,{\rm GeV}^2$ at 
$\sqrt s = 314\,{\rm GeV}$ in comparison with the H1~[39] data. 
Curves {\sl 1 --- 3} are the same as in Fig.~3.

\vspace{1cm}
\noindent
{\bf Fig.~5.} The double differential cross sections of the inelastic 
$J/\psi$ leptoproduction obtained in the kinematical region 
$2 < Q^2 < 100\,{\rm GeV}^2$, $50 < W < 225 \,{\rm GeV}$, 
$0.3 < z < 0.9$ and ${\bf p}_{\psi\,T}^{*\,2} > 1\,{\rm GeV}^2$ at 
$\sqrt s = 314\,{\rm GeV}$ in comparison with the H1~[39] data. 
Curves {\sl 1 --- 3} are the same as in Fig.~3.

\vspace{1cm}
\noindent
{\bf Fig.~6.} The total and single differential cross sections of the 
inelastic $J/\psi$ photoproduction obtained in the kinematical region 
$60 < W < 240 \,{\rm GeV}$, $1 < {\bf p}_{\psi\,T}^{2} < 60\,{\rm GeV}^2$,
$0.3 < z < 0.9$ in comparison with the H1~[38] data. 
Curves {\sl 1 --- 3} are the same as in Fig.~3.

\vspace{1cm}
\noindent
{\bf Fig.~7.} The total and single differential cross sections of the 
inelastic $J/\psi$ photoproduction obtained in the kinematical region 
$50 < W < 180 \,{\rm GeV}$, ${\bf p}_{\psi\,T}^{2} > 1\,{\rm GeV}^2$,
$0.4 < z < 0.9$ in comparison with the ZEUS~[40] data. 
Curves {\sl 1 --- 3} are the same as in Fig.~3.

\vspace{1cm}
\noindent
{\bf Fig.~8.} The double differential cross sections of the 
inelastic $J/\psi$ photoproduction obtained in the kinematical region 
$60 < W < 240 \,{\rm GeV}$, $1 < {\bf p}_{\psi\,T}^{2} < 60\,{\rm GeV}^2$,
$0.3 < z < 0.9$ in comparison with the H1~[38] data. 
Curves {\sl 1 --- 3} are the same as in Fig.~3.

\vspace{1cm}
\noindent
{\bf Fig.~9.} The parameter $\alpha$ as a function $z$ and 
${\bf p}_{\psi\,T}$ for the inelastic $J/\psi$ photoproduction process 
which obtained in the kinematical region $60 < W < 240 \,{\rm GeV}$, 
$0.3 < z < 0.9$ and $1 < {\bf p}_{\psi\,T}^{2} < 60\,{\rm GeV}^2$
in comparison with the H1~[38] data.
Curve {\sl 1} corresponds to the SPM calculations 
at the leading order approximation with GRV (LO) gluon density, 
curve {\sl 2} corresponds to the $k_T$-factorization QCD 
calculations with JB unintegrated gluon distribution.

\vspace{1cm}
\noindent
{\bf Fig.~10.} The parameter $\alpha$ as a function ${\bf p}_{\psi\,T}$ 
for the inelastic $J/\psi$ photoproduction process 
which obtained in the kinematical region $50 < W < 180 \,{\rm GeV}$, 
$0.4 < z < 0.9$ (Fig.~10a, c), $0.4 < z < 1$ (Fig.~10b, d) and 
$1 < {\bf p}_{\psi\,T}^{2} < 60\,{\rm GeV}^2$ in comparison with 
the ZEUS~[40] data. Curves {\sl 1} and {\sl 2} are the same as in Fig.~9.

\end{large}
\end{document}